\documentclass[12pt]{article}
\usepackage{graphics,amssymb,latexsym,amsmath,epsfig,color,graphicx,wrapfig,multicol}

\newcommand\pubnumber{FERMILAB-CONF-15-342-T}
\newcommand\pubdate{\today}

\def\fermilab{Fermi National Accelerator Laboratory\\ P.O. Box 500, Batavia, Illinois 60510-5011, USA}

\def\Title#1{\begin{center} {\Large #1 } \end{center}}
\def\Author#1{\begin{center}{ \sc #1} \end{center}}
\def\Address#1{\begin{center}{ \it #1} \end{center}}

\newcommand\pubblock{\rightline{\begin{tabular}{l} \pubnumber\\
         \pubdate  \end{tabular}}}
\newenvironment{Abstract}{\begin{quotation}  }{\end{quotation}}
\newenvironment{Presented}{\begin{quotation} \begin{center} 
             PRESENTED AT\end{center}\bigskip 
      \begin{center}\begin{large}}{\end{large}\end{center} \end{quotation}}
\def\Acknowledgements{\bigskip  \bigskip \begin{center} \begin{large}
             \bf ACKNOWLEDGEMENTS \end{large}\end{center}}

\def\beq{\begin{equation}}
\def\eeq#1{\label{#1}\end{equation}}
\def\eeqn{\end{equation}}

\def\beqa{\begin{eqnarray}}
\def\eeqa#1{\label{#1}\end{eqnarray}}
\def\eeqan{\end{eqnarray}}

\let\bar=\overbar

\def\Dslash{\not{\hbox{\kern-4pt $D$}}}
\def\dslash{\not{\hbox{\kern-2pt $\del$}}}

\def\msb{{\bar{\ssstyle M \kern -1pt S}}}

\begin{document}
\begin{titlepage}
\pubblock

\vfill
\Title{Recent progress in lattice calculations of properties of open-charm mesons}
\vfill
\Author{Daniel Mohler}
\Address{\fermilab}
\vfill
\begin{Abstract}
Recent progress in lattice calculations of properties of open-charm mesons, both regular and exotic, is reviewed, with an emphasis on spectroscopy. After reviewing recent calculations of excited state energy levels I will discuss progress in extracting hadronic masses and widths of charmed states from Lattice QCD simulations including low-lying scattering channels directly, to determine phase shift data and bound state/ resonance properties. With regard to other properties results from recent calculations of the $DD^*\pi$ and $DD\rho$, $D^*D^*\rho$ couplings are presented. Beyond regular mesons, searches for explicitly exotic (tetraquark) states are also reviewed.

\end{Abstract}
\vfill
\begin{Presented}
The 7th International Workshop on Charm Physics (CHARM 2015)\\
Detroit, MI, 18-22 May, 2015
\end{Presented}
\vfill
\end{titlepage}
\def\thefootnote{\fnsymbol{footnote}}
\setcounter{footnote}{0}
%

\section{Introduction}

Lattice QCD, the simulation of Quantum Chromodynamics (QCD) on a space-time grid, is commonly regarded as the ideal tool to investigate the non-perturbative physics of the strong interaction at hadronic energy scales. Indeed there has been impressive progress in recent years, with simulations at or close to physical quark masses with dynamical light, strange and even charm quarks. For ground state observables involving mesons only, many quantities have been calculated with both full control of the relevant systematic uncertainties and with an impressive statistical and systematic precision. The masses of a number of low-lying hadrons have been determined in this way \cite{Kronfeld:2012uk,Dowdall:2012ab} including ground states with charmed quarks. Example of quantities determined with full control of systematic uncertainties are collected in the FLAG review \cite{Aoki:2013ldr}. The review collects lattice results relevant for flavor physics and tries to provide the current best lattice values for particular quantities listing the various lattice results categorized by the number of dynamic flavors and rating them by fixed quality criteria.

For properties of hadronic excitations or for hadrons close to multi-hadron thresholds the situation is however complicated, and fully systematic determinations of excited state properties are still lacking. The reason for this is that excited states are much more challenging, both conceptually and computationally:

\begin{itemize}
\item Looking at Euclidean-space correlation functions, excited state contributions appear as subleading exponentials
\begin{align*}
\left <\hat{O}_2(t)\hat{O}_1(0)\right >_T &\propto\sum_{n}e^{-t{E_n}}<0|\hat{O}_2|n><n|\hat{O}_1|0>\; ,
\end{align*}
where $E_n$ is the energy of the n-th state. For a given channel the sum in principle consists of all hadronic (including multi-hadron) states with the same quantum numbers. This makes the reliable extraction of excited state energy levels in Monte-Carlo simulations with non-vanishing statistical uncertainty very challenging.
\item For hadronic resonances and close-to-threshold bound states, scattering phase-shifts and consequently resonance and bound state poles are not directly accessible in Euclidean space \cite{Maiani:1990ca}, but instead have to be extracted using the finite volume method pioneered by L\"uscher \cite{Luscher:1985dn,Luscher:1986pf,Luscher:1990ux}. 
\item In determining energy levels from lattice calculations using quark-antiquark interpolating fields for mesons and 3-quark interpolating fields for baryons, it became obvious \cite{McNeile:2002fh,Engel:2010my,Bulava:2010yg,Dudek:2010wm} that a basis of only quark-antiquark or only scattering operators is not very useful and is probably not suitable to obtain the true energy levels. Modern calculations therefore employ a mixed basis of both scattering and single-hadron interpolating fields, which for many interesting quantum numbers necessitates the inclusion of quark-line diagrams with backtracking quark lines, (inspired by single hadrons) often referred to as ``disconnected diagrams''. These are especially challenging numerically.
\end{itemize}

Consequently most calculations of charmed excitations are of an exploratory nature and currently lack a detailed estimate of the systematic uncertainty in the calculation. A number of results reported in this review are also preliminary in the sense that they have not yet been published.

\section{Single hadron spectra}

\begin{figure}[tb]
\centering
\includegraphics[clip,width=7.5cm]{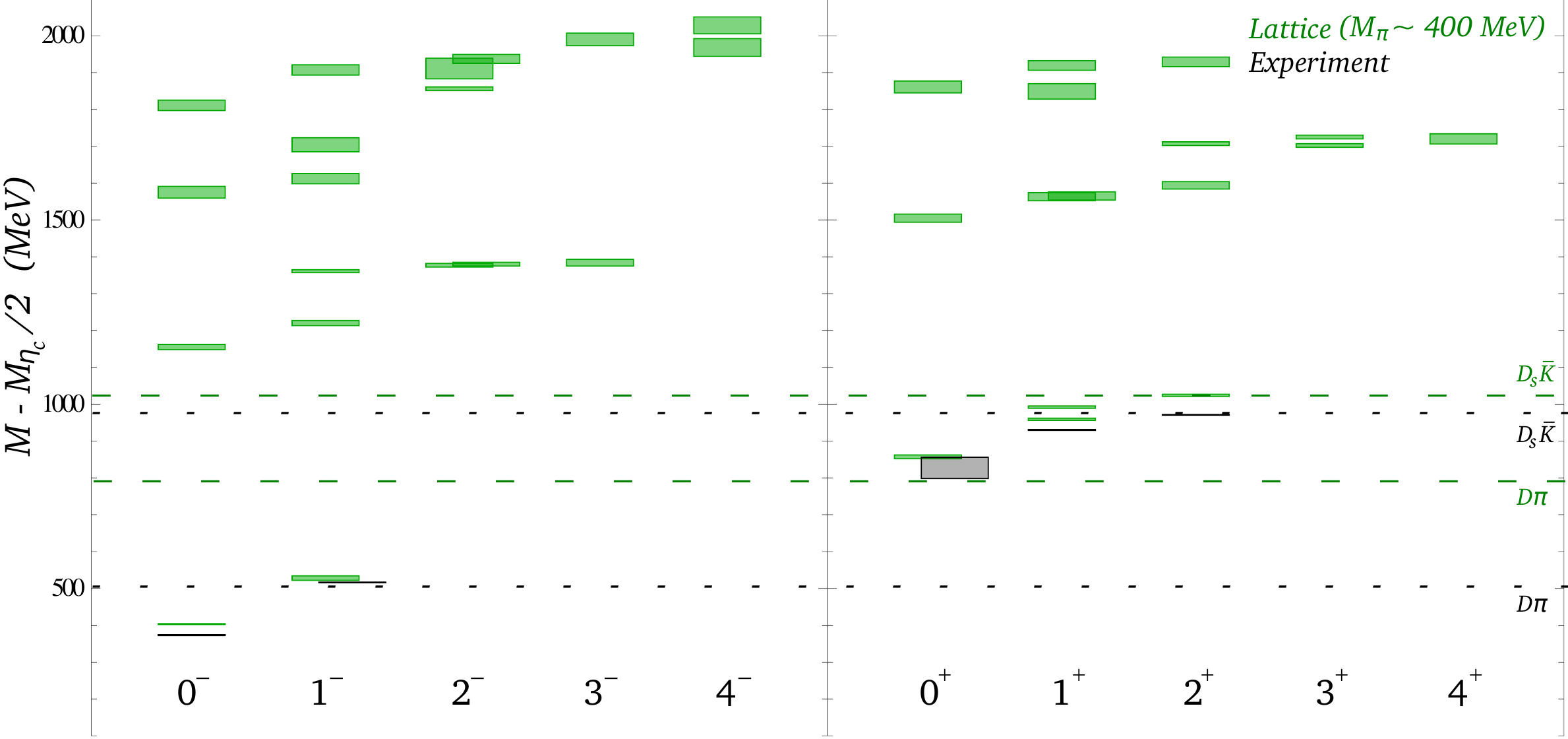}
\includegraphics[clip,width=7.5cm]{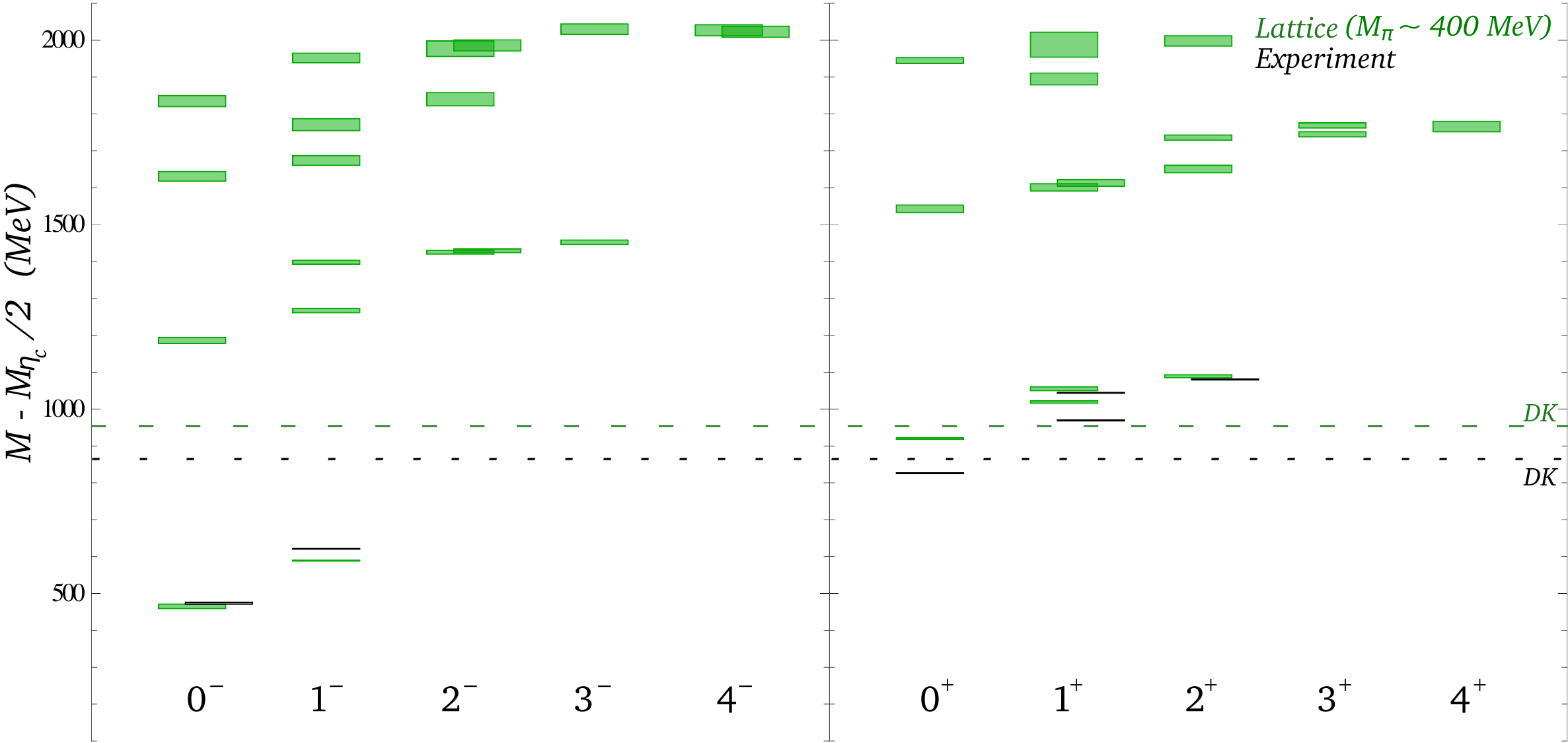}
\caption{$D$ and $D_s$ meson spectra by the Hadron Spectrum Collaboration \cite{Moir:2013ub}. The $J^P$ quantum numbers are denoted on the horizontal axis. All results are for up and down quark masses corresponding to 400MeV pions. Lattice data in green compared to mesons known from experiment in black.}
\label{fig:HSC_heavylights}
\end{figure}

As outlined in the previous section, one of the challenges for lattice calculations of excited state masses consists of reliably determining excited state energy levels. Figure \ref{fig:HSC_heavylights} shows the energy levels determined by the Hadron Spectrum Collaboration from single meson interpolating fields for charmed and charmed-strange mesons \cite{Moir:2013ub}. The impressive results demonstrate that a large number of energy levels in all low-lying $J^P$ quantum number channels can be determined. The authors use information from the overlaps to certain interpolating fields to reliably identify the spin \cite{Dudek:2009qf}, which is challenging due to the rotational symmetry breaking on the lattice. In addition to the single-hadron energy levels, the lowest scattering thresholds in the system are illustrated by the dashed lines. Above these thresholds, the relation of the observed energy levels to hadronic resonances is not straightforward. It should also be pointed out that energy levels that arise purely from scattering thresholds can not be distinguished from meson excitations when using only single-hadron interpolators. However no clear candidates for such energy levels are seen in their simulation, suggesting that the number of energy levels observed can be related to mesons with the given quantum numbers.

\begin{wrapfigure}{r}{0.6\textwidth}
\centering
\includegraphics[clip,width=0.6\textwidth]{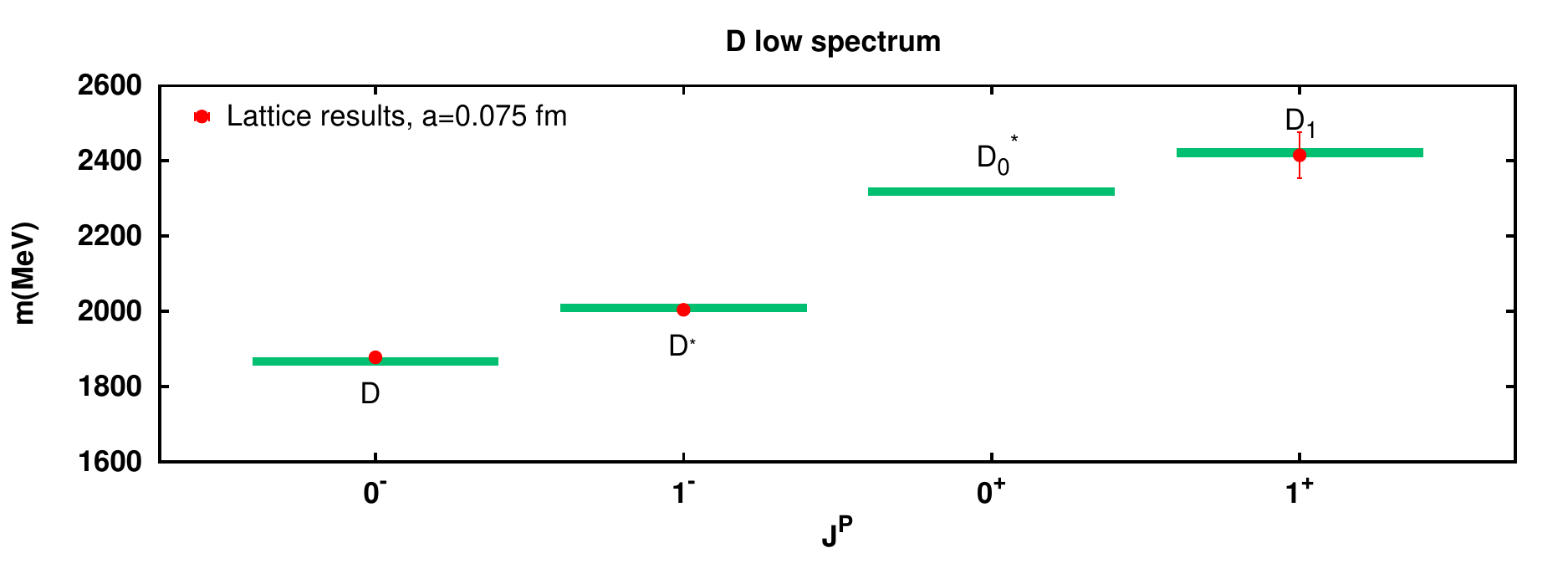}\\
\includegraphics[clip,width=0.6\textwidth]{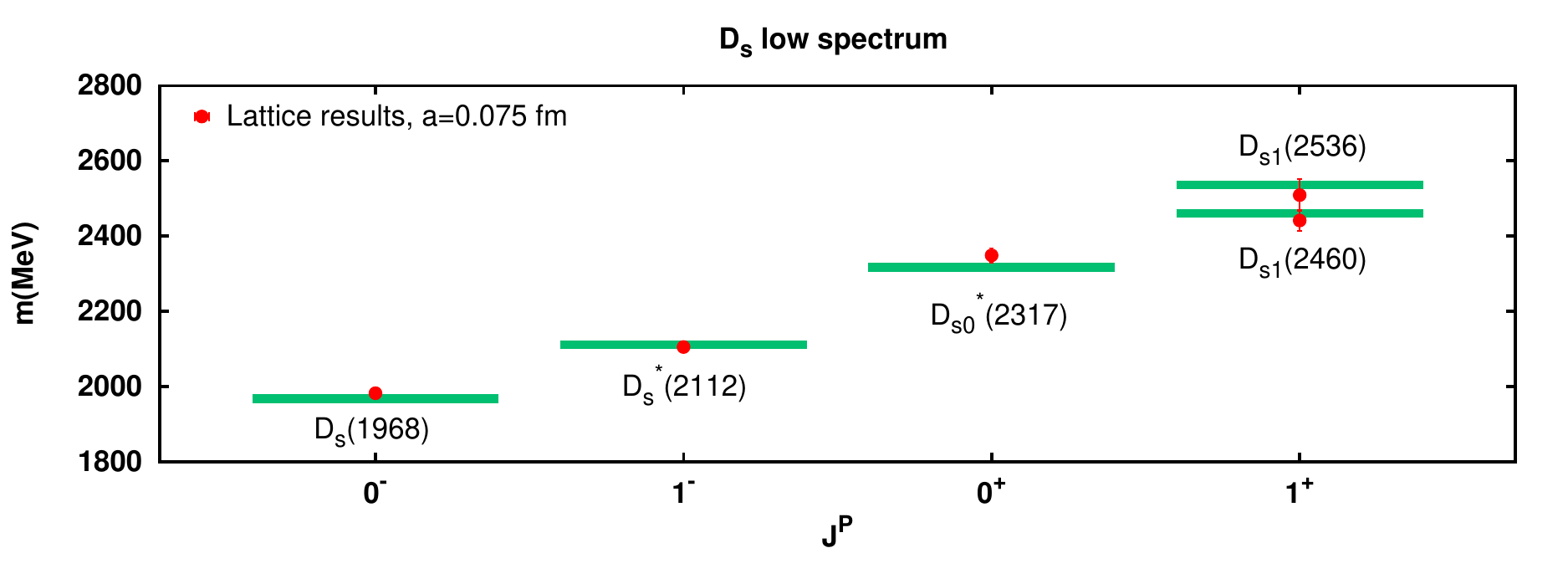}
\caption{Low-lying $D$ (top) and $D_s$ (bottom) meson spectra from \cite{Bali:2015lka}.}
\label{fig:Regensburg}
\end{wrapfigure}

Figure \ref{fig:Regensburg} shows low-lying positive parity charmed-states calculated as part of a study of charmed baryons by the QCDSF collaboration \cite{Bali:2015lka}. They calculated the low-lying spectra on ensembles with 2+1 flavors of dynamical quarks (for 2 volumes and 3 pion masses with $259\mathrm{MeV} \le m_\pi\le 460$MeV) and the figure shows results at a lattice spacing of 0.075fm. For the $D$ mesons they observe ground states corresponding to $D\pi$ and $D^*\pi$ for $J^P=0^+$ and $1^+$ which are not shown in the figure. Their basis was not large enough to observe a second $D_1$ state. For charmed-strange states the authors comment that the $D_{s1}(2536)$ state might be the $D^*K$ scattering state. For the $D_{s0}^*$ with $J^P=0^+$ their analysis uses a only single interpolator and the presence of a signal due to the $DK$ scattering threshold can not be excluded. In combination the results presented here nicely illustrate the pitfalls of using only single-hadron interpolators: The observed overlaps to scattering states seem to depend strongly on the technicalities used in the study. This is clearly not a satisfactory situation.
\begin{wrapfigure}[14]{r}{0.6\textwidth}
\centering
\includegraphics[clip,width=0.6\textwidth]{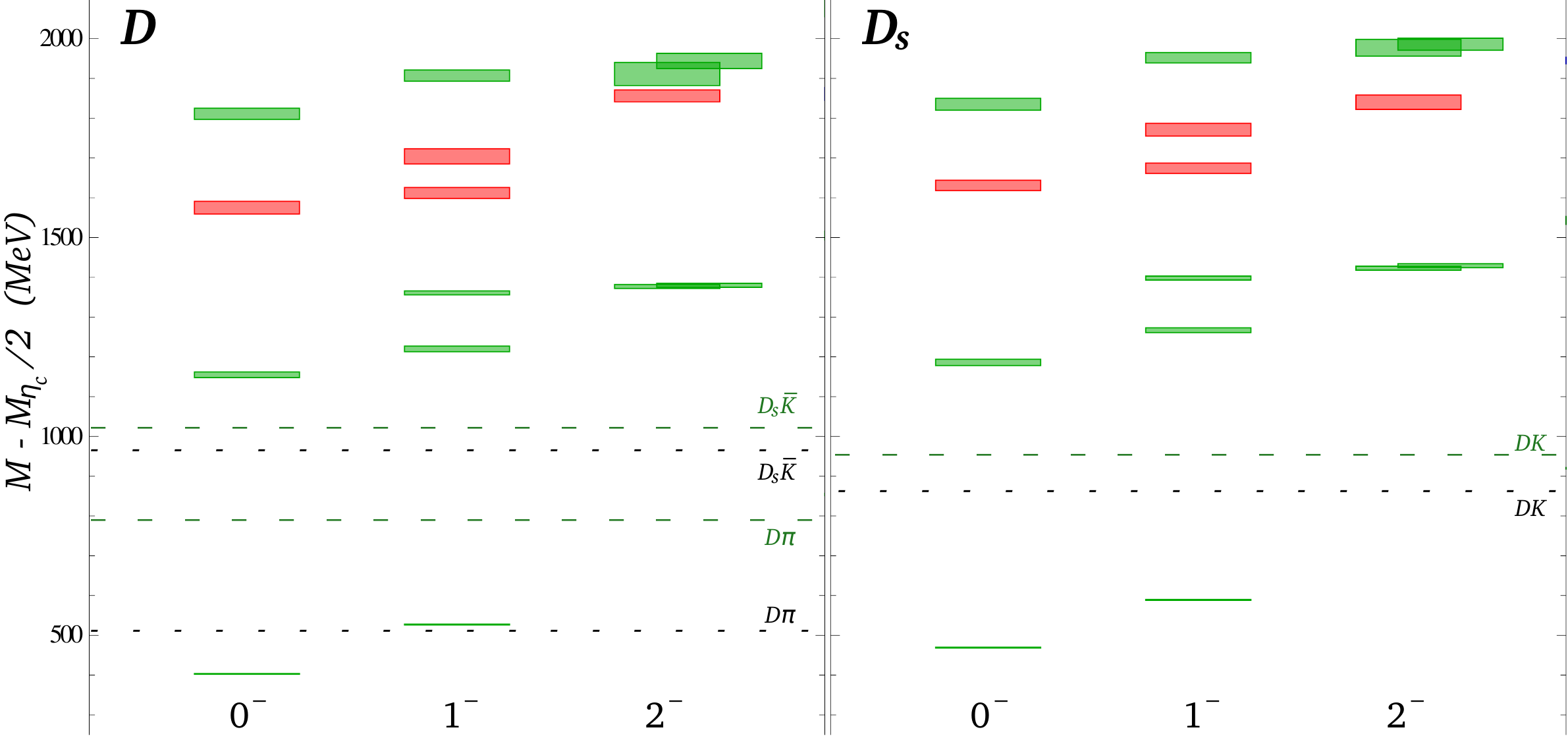}
\caption{Spectrum of $D$ and $D_s$ meson energy levels of negative parity from \cite{Moir:2013ub} with hybrid meson energy levels shown in red.}
\label{fig:HSC_hybrids}
\end{wrapfigure}

While in general the relation of excited state energy levels to resonances is not straightforward, Figure \ref{fig:HSC_hybrids} highlights another feature from the HSC calculation of charmed mesons \cite{Moir:2013ub}: Their data clearly reveals hybrid meson candidates (containing both valence quarks and excited glue) with negative parity. Such states have been conjectured and are seen in the lattice calculation for various quantum numbers.

\section{Charmed mesons and scattering}

\begin{wrapfigure}[17]{r}{0.5\textwidth}
\centering
\includegraphics[clip,width=0.5\textwidth]{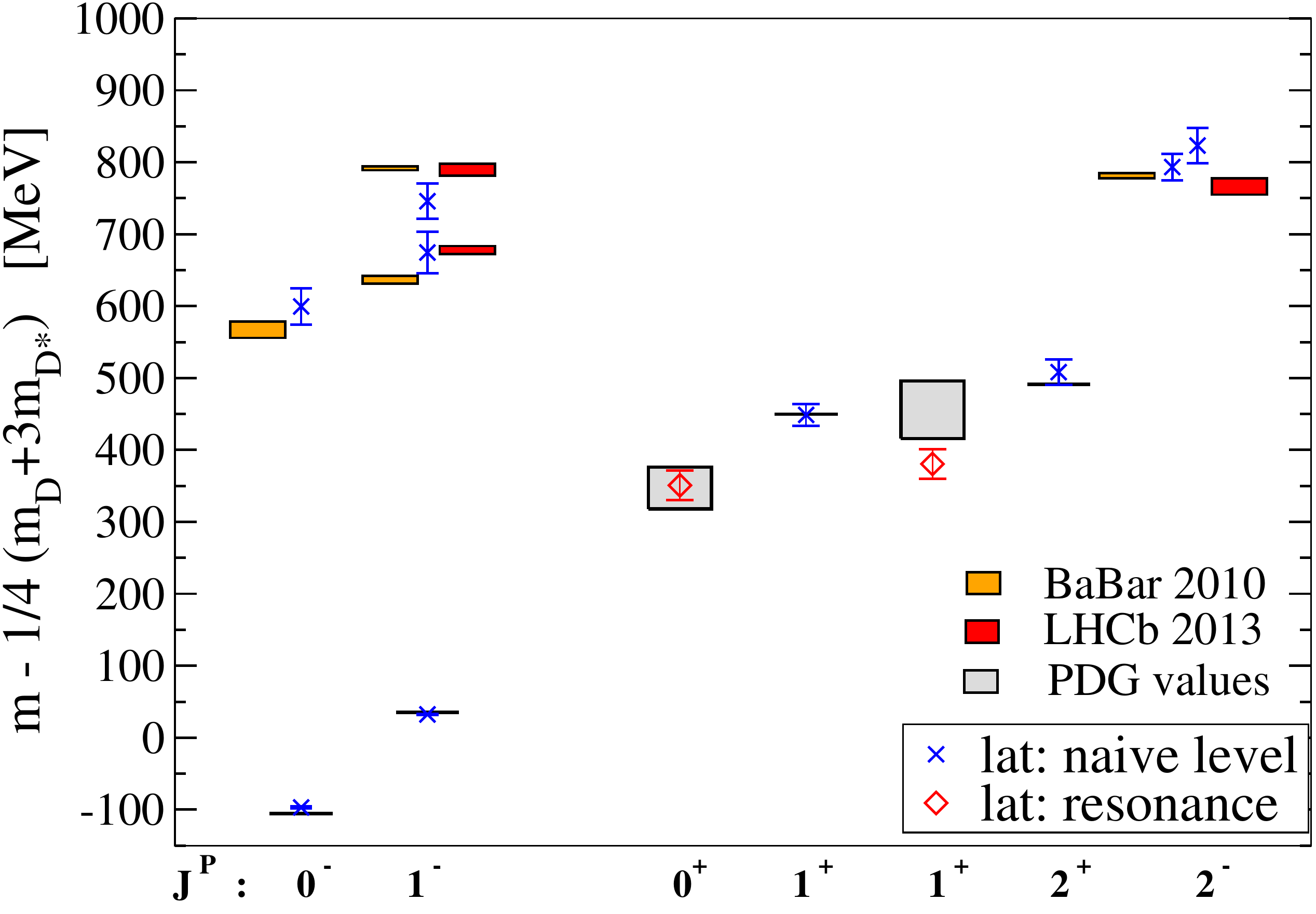}
\caption{Low lying $D$ meson states from \cite{Mohler:2012na} compared to experiment. Only the red diamonds are resonance masses, while all other states are naive energy levels.}
\label{fig:Dmesons}
\end{wrapfigure}

To determine scattering phase shifts and the masses and widths of resonances or the pole positions of bound states from Lattice QCD, L\"uscher's finite volume method \cite{Luscher:1985dn,Luscher:1986pf,Luscher:1990ux} and its various extensions (for a review see \cite{Briceno:2014pka}) can be used. Within this framework 2-hadron scattering and transitions are well understood and there is some progress for 3 (or more) hadrons. Just as described in the last section, the determination of energy levels is the first step. L\"uscher's formula and its extensions then relate the energy shifts in finite volume with respect to scattering thresholds to the phase shift(s) of the continuum scattering amplitude.

\begin{wraptable}[8]{r}{0.5\textwidth}
\centering
\begin{tabular}{ccc}
\hline
&$D_0^*(2400)$&$D_1(2430)$\\
\hline
$g^{lat}$ [GeV] & $2.55 \pm 0.21$ & $2.01 \pm 0.15$\\
$g^{exp}$ [GeV] & $1.92 \pm 0.14$ & $2.50 \pm 0.40$\\
\hline
\end{tabular}
\caption{Results for the couplings g from \cite{Mohler:2012na}}
\label{tab:Dmesons}
\end{wraptable}

For charmed mesons this method has first been applied by taking a look at the lowest positive parity $D$ meson resonances in s-wave $D\pi$ and $D^*\pi$ scattering \cite{Mohler:2012na}. The exploratory study on lattices with $m_\pi=266$MeV determined the couplings $g$ rather than $\Gamma=g^2\frac{p^*}{s}$. Figure \ref{fig:Dmesons} shows the resulting resonance masses 
\begin{wrapfigure}[14]{r}{0.5\textwidth}
\centering
\includegraphics[clip,width=0.5\textwidth]{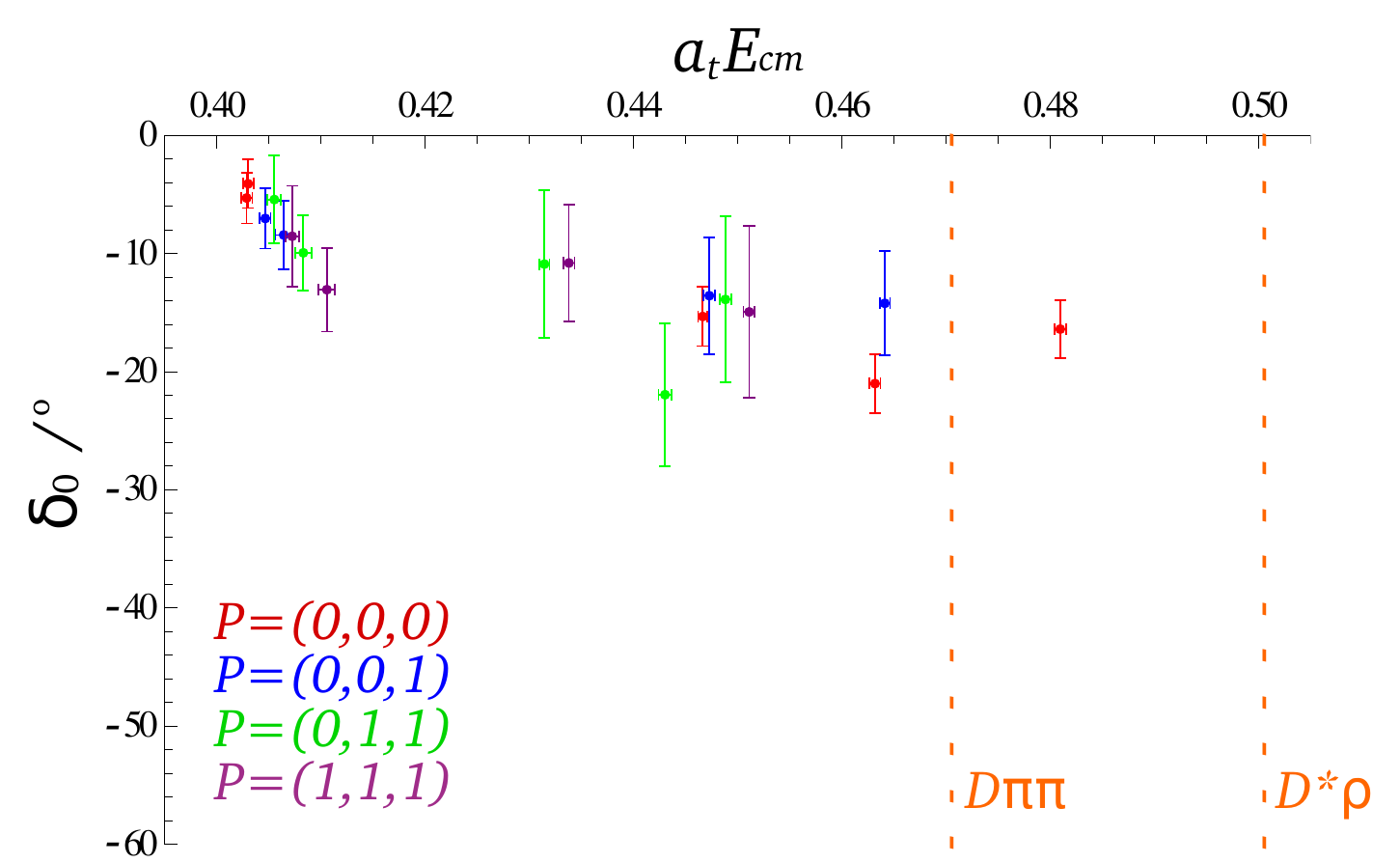}
\caption{Preliminary results for the $D\pi$ Isospin $I=\frac{3}{2}$ phase shift from \cite{Moir:2013yfa} considering the lowest partial wave only.}
\label{fig:HSC_phaseshift}
\end{wrapfigure}

\noindent along with naive energy levels for some other D-meson states. Table \ref{tab:Dmesons} shows the resulting couplings, which are in qualitative agreement with experiment.

\begin{figure}[tb]
\centering
\includegraphics[clip,width=7.3cm]{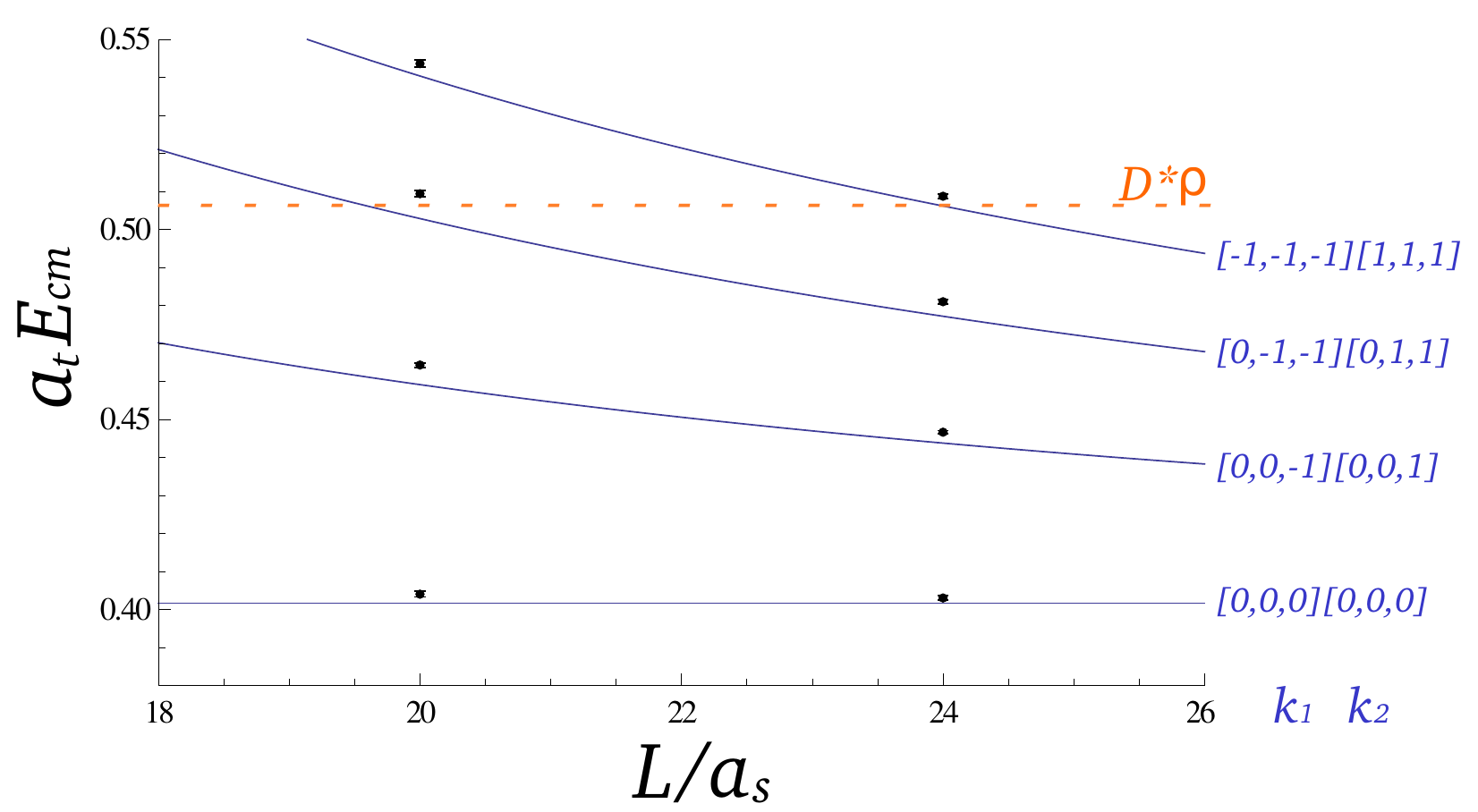}
\includegraphics[clip,width=7.3cm]{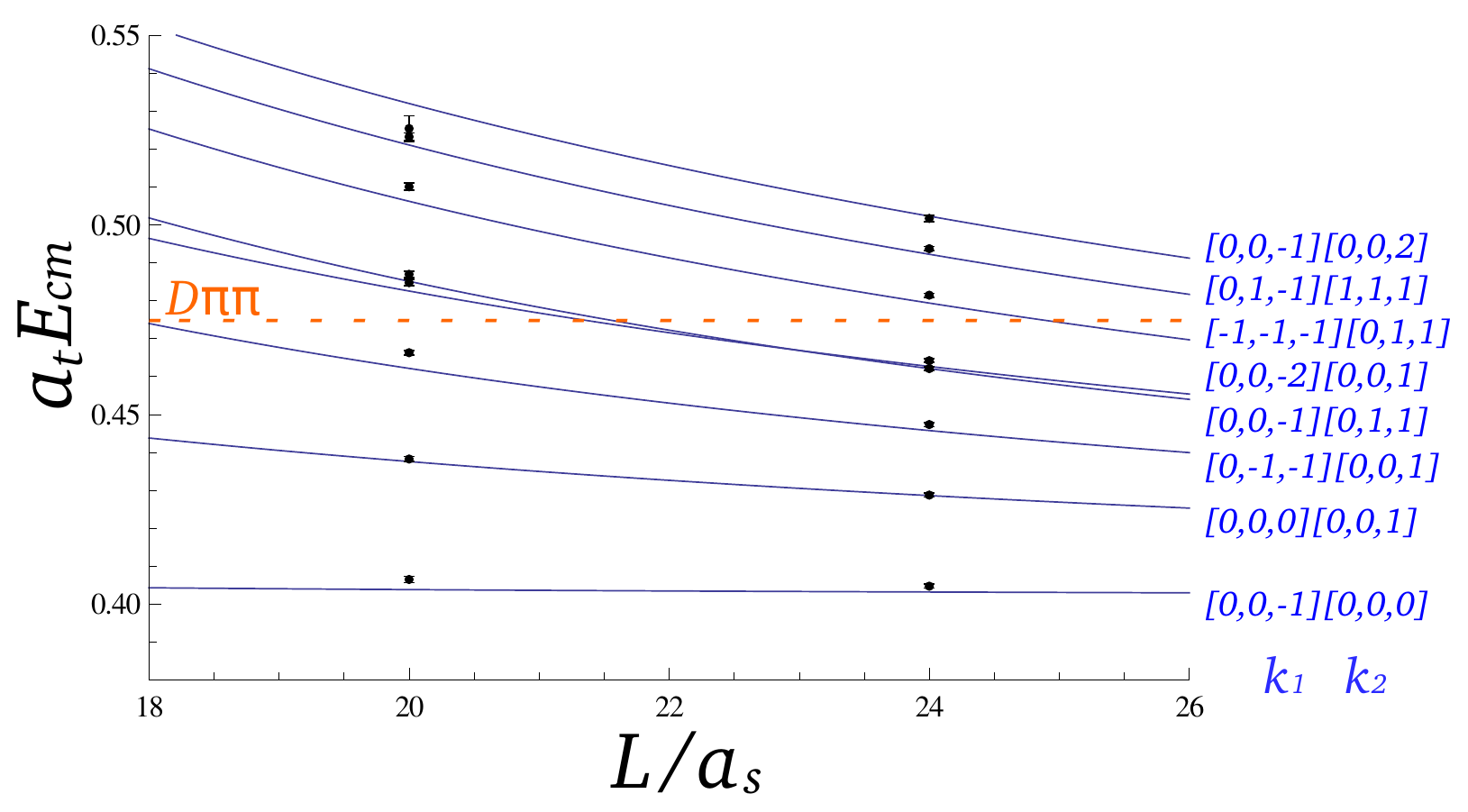}\\
\includegraphics[clip,width=7.3cm]{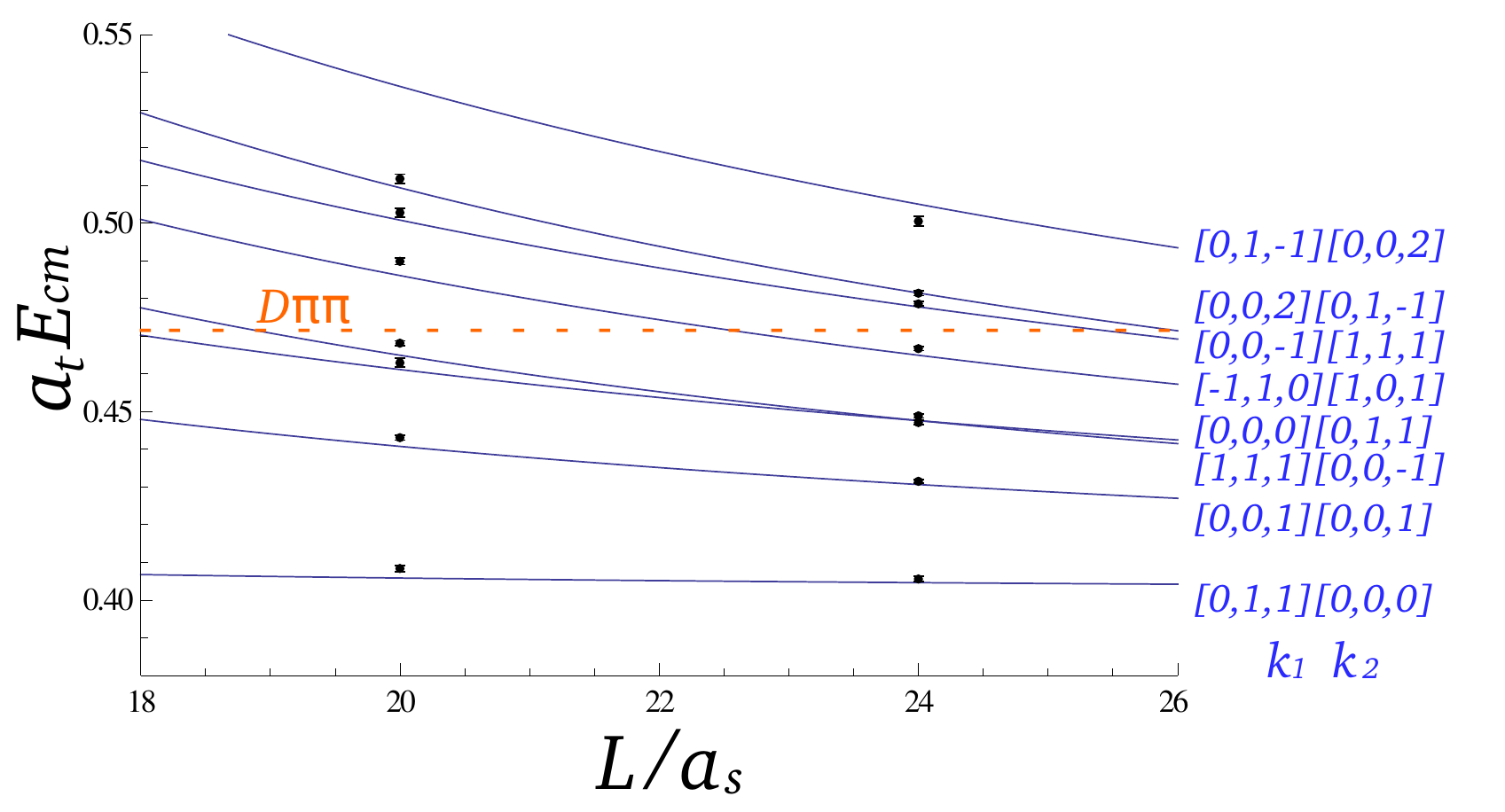}
\includegraphics[clip,width=7.3cm]{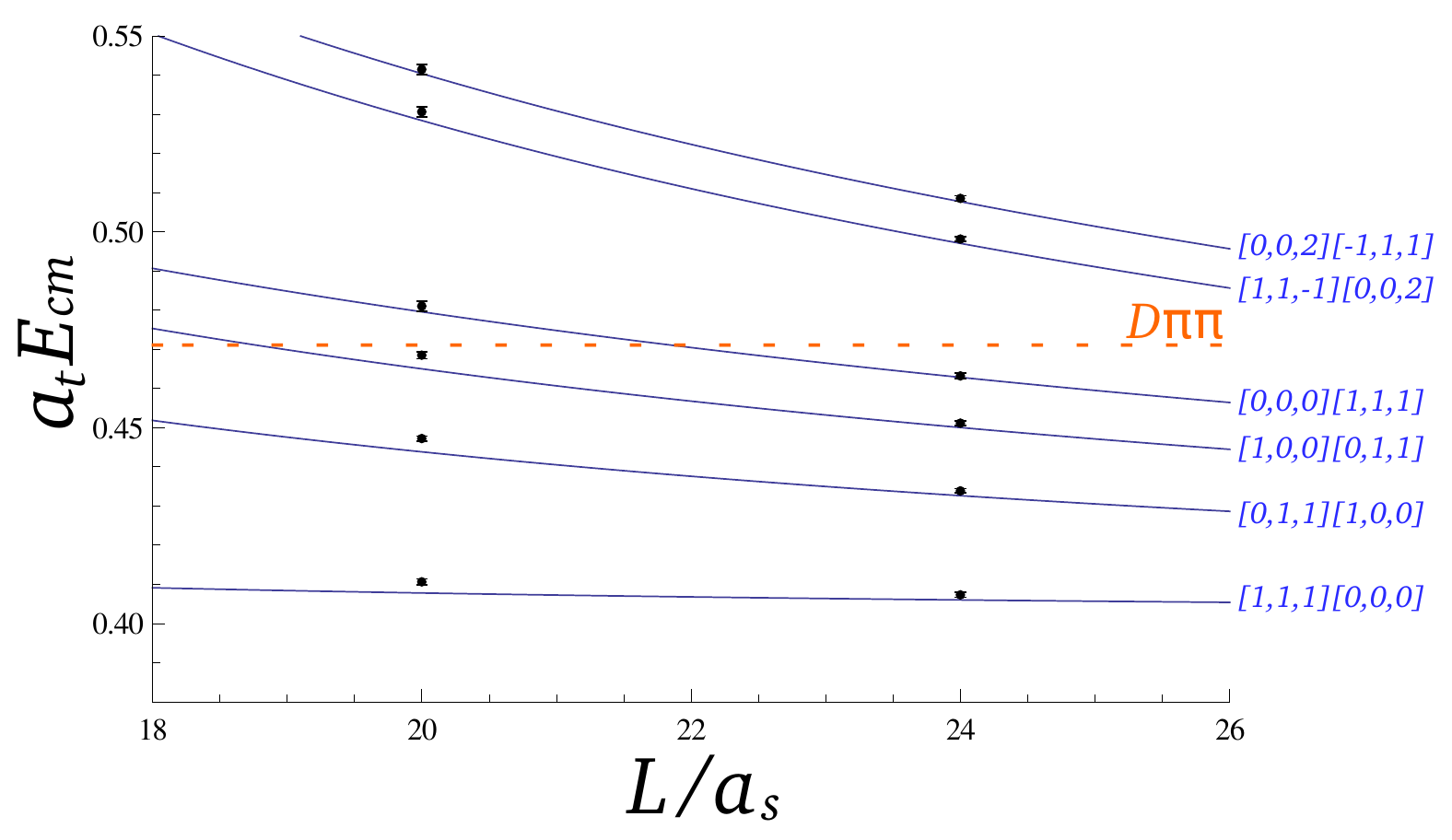}
\caption{Preliminary results for energy levels of $D\pi$-scattering Isospin $I=\frac{3}{2}$ from the Hadron Spectrum Collaboration \cite{Moir:2013yfa}.}
\label{fig:HSC_spectrum}
\end{figure}

As demonstrated by preliminary results from the Hadron Spectrum Collaboration \cite{Moir:2013yfa}, these results can be improved by considering several volumes and multiple moving frames. Preliminary results for $D\pi$-scattering Isospin $I=\frac{3}{2}$ are shown in Figs. \ref{fig:HSC_phaseshift} and \ref{fig:HSC_spectrum}. Figure \ref{fig:HSC_spectrum} shows the energies compared to the non-interacting energy levels for multiple momentum frames. Figure \ref{fig:HSC_phaseshift} shows the extracted phase shifts considering only the lowest partial wave for this preliminary analysis. A dense coverage of interesting energy region is obtained.

\begin{figure}[h!]
\centering
\includegraphics[clip,width=7.5cm]{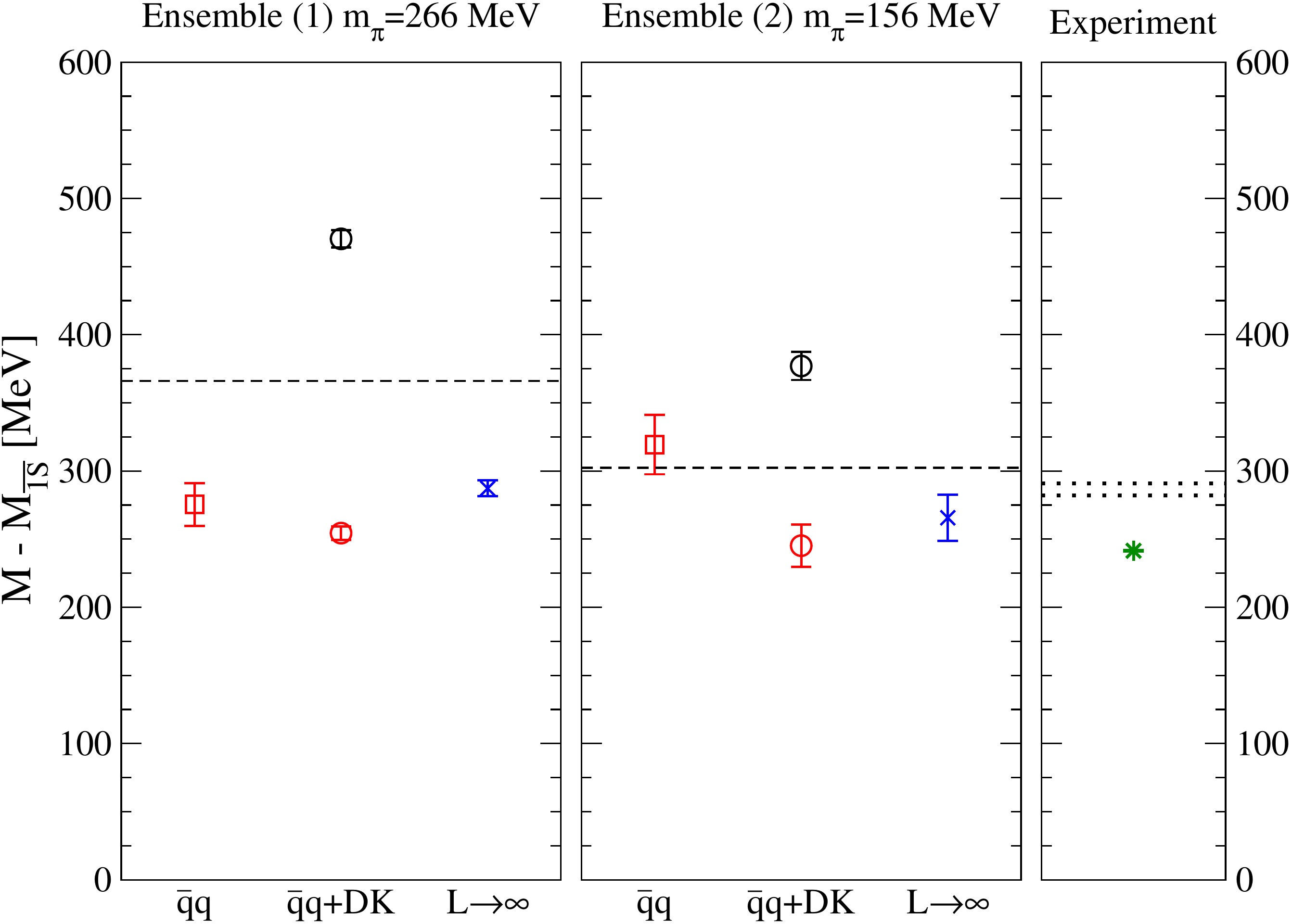}
\includegraphics[clip,width=7.5cm]{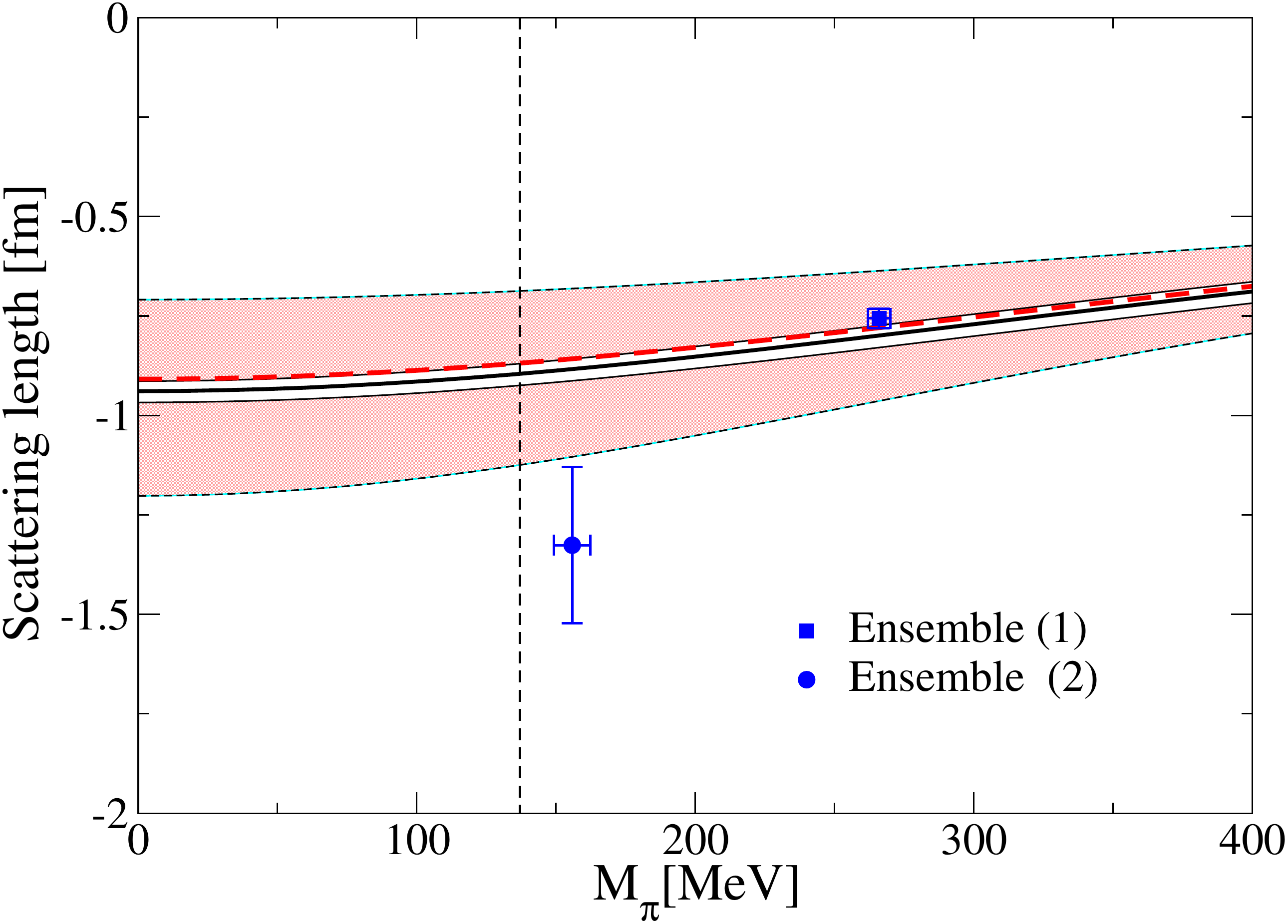}
\caption{Left: Energies and pole positions for the $D_{s0}^*(2317)$ bound state from \cite{Mohler:2013rwa} at two different pion masses compared to experiment. Right: Comparison of the s-wave scattering length to predictions from an indirect calculation \cite{Liu:2012zya}.}
\label{fig:ds2317_energies}
\end{figure}

For charmed-strange p-wave mesons recent results \cite{Mohler:2013rwa,Lang:2014yfa} including the $DK$ and $D^*K$ scattering states explicitly into the basis along with quark-antiquark interpolators for determinations of the $D_{s0}^*(2317)$ and $D_{s1}(2460)$ resolve a long standing puzzle. A much better quality of the ground state plateau is observed with a combined basis and unlike the quark model predictions and older lattice results the results agree qualitatively with experiment (see Figure \ref{fig:ds2317_energies}). The resulting S-wave scattering length agrees with an indirect determination from EFT-calculations with low-energy constants determined by lattice simulations in different scattering channels \cite{Liu:2012zya}, as displayed in Figure \ref{fig:ds2317_energies}.

In \cite{Torres:2014vna} a reanalysis of the lattice data from \cite{Mohler:2013rwa,Lang:2014yfa} is performed by using an auxiliary potential $V$ and extracting the parameters of $V$ from the lattice data.
\begin{align*}
V&=\alpha+\beta(s-s_{th}) \qquad \tilde{T}=\frac{1}{V^{-1}-\tilde{G}}\; ,\\
\tilde{G}&=G+\lim_{q_{max}\rightarrow\infty}\left(\frac{1}{L^3}\sum_{q_i}^{q_{max}}I(\vec{q_i})-\int_{q<q_{max}}\frac{d^3q}{(2\pi)^3}I(\vec{q})\right)\; .
\end{align*}
A generalization of Weinberg's compositeness condition then yields the probability of finding $DK$ ($D^*K$) components in the physical bound states
\begin{align*}
P(KD)=0.76(12) \quad\mbox{for the}\quad D_{s0}^*(2317)\;,\\
P(KD^*)=0.53(17) \quad\mbox{for the}\quad D_{s1}(2460)\; .
\end{align*}
\begin{wrapfigure}[21]{r}{0.47\textwidth}
\centering
\includegraphics[width=0.45\textwidth,clip]{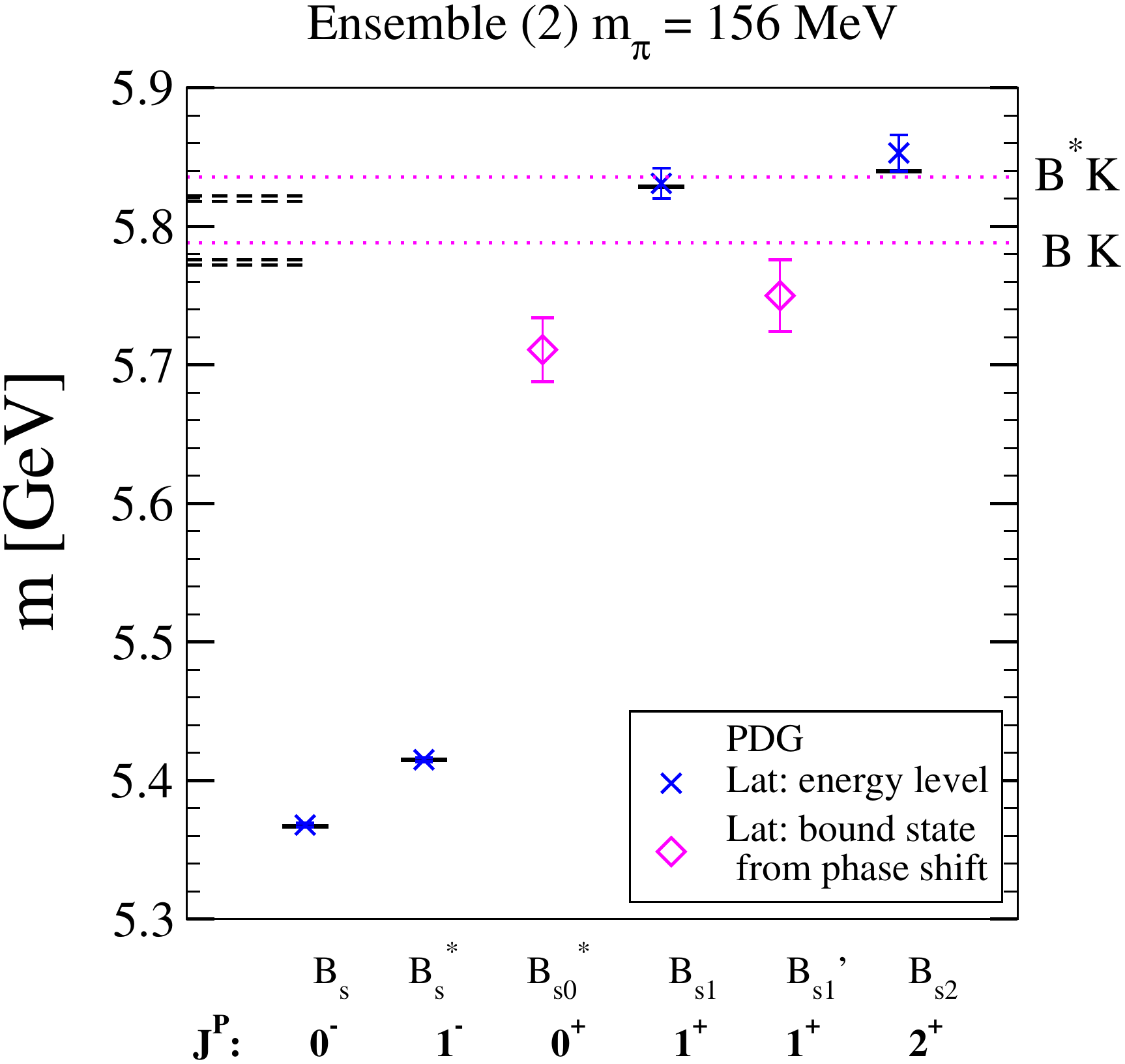}
\caption{Prediction of the $B_s$ spectrum analogues of the $D_{s0}^*(2317)$ and the $D_{s1}^*(2460)$ from \cite{Lang:2015hza}. The uncertainty estimate for the bound state poles is shown by adding statistical and systematic uncertainties in quadrature.}
\label{fig:heavystrange_overview}
\end{wrapfigure}
The authors further suggest a study with increased precision which should also include the $D^{(*)}\eta$ scattering channel. As a word of caution, it should also be mentioned that the lattice data used suffers from sizable discretization effects, so any results about the composition derived from the data has to be taken with a grain of salt.

Figure \ref{fig:heavystrange_overview} shows a prediction of the related $j=\frac{1}{2}$ states with $J^P=0^+,1^+$ in the bottom-strange meson spectrum \cite{Lang:2015hza}. These states have not yet been observed in experiment.

\begin{figure}[tbh]
\centering
\includegraphics[clip,width=7.5cm]{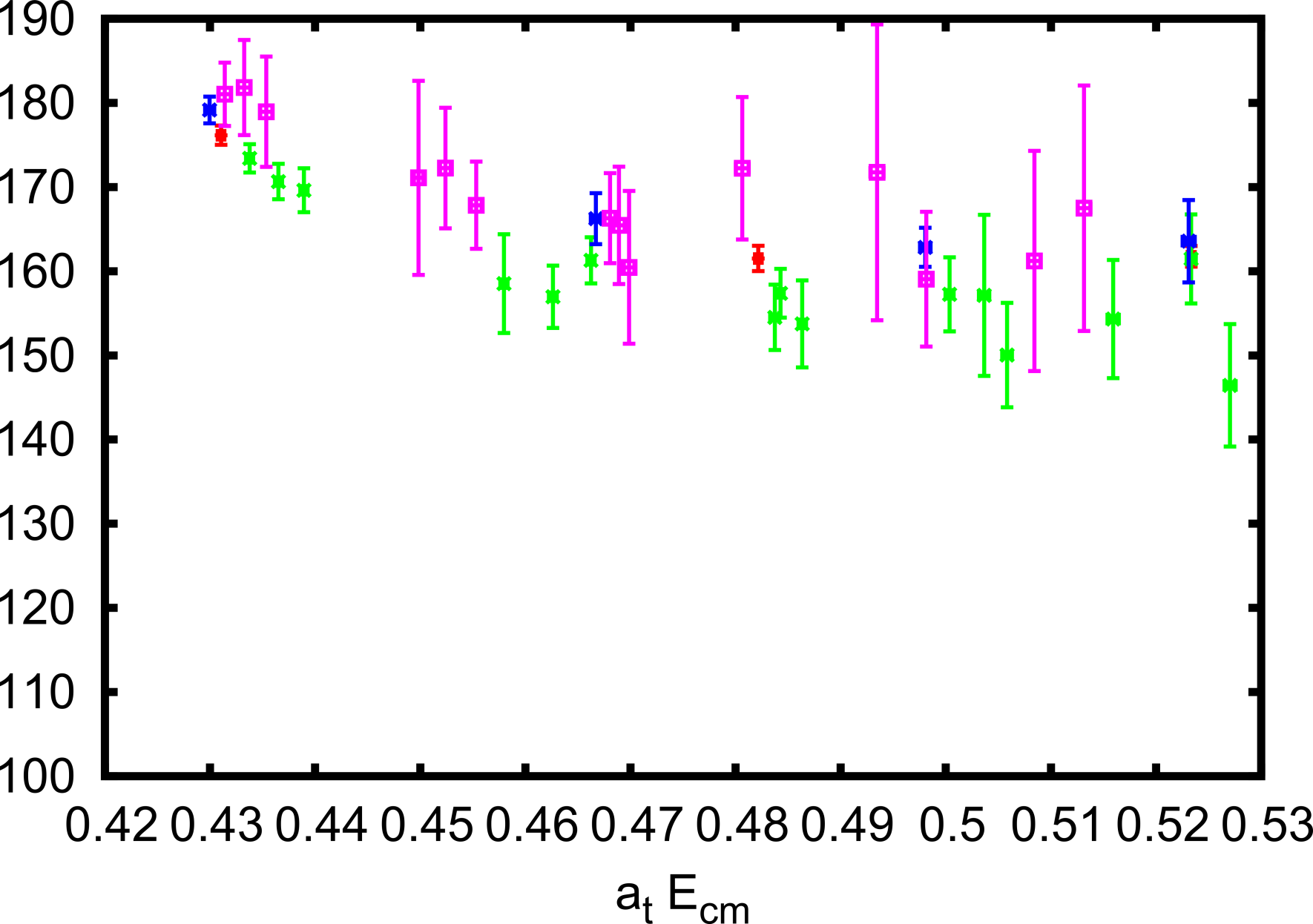}
\caption{Preliminary results for the $DK$ s-wave scattering phase shift in Isospin $I=1$ from \cite{Moir:2013yfa} considering the lowest partial wave only.}
\label{fig:HSC_phaseshift2}
\end{figure}

Further interesting results for the scattering of Kaons with D-mesons can be expected from the Hadron Spectrum Collaboration. Figure \ref{fig:HSC_phaseshift2} shows preliminary results for $DK$ s-wave scattering in Isospin 1 \cite{Moir:2013yfa}, considering only the lowest partial wave. Similar results for quantum numbers with resonances can be expected in the future.

\section{Calculations of $g_{DD^*\pi}$ and $g_{DD\rho}$, $g_{D^*D^*\rho}$}

Beyond masses, there is also progress in calculating further properties of charmed mesons. A recent example is the calculation of the $DD^*\pi$ and $DD\rho$, $D^*D^*\rho$ couplings from (transition) matrix elements

\begin{align*}
&<D(p^\prime)|A_\mu(q)|D^*(p,s)>\quad\mbox{with}\quad A_\mu=\bar{u}\gamma_5\gamma_id\; ,\\
&<D(p^\prime)|V_\mu(q)|D(p)>\quad\mbox{with}\quad
V_\mu=\frac{2}{3}\bar{c}\gamma_\mu c+\frac{2}{3}\bar{u}\gamma_\mu
u-\frac{1}{3}\bar{d}\gamma_\mu d\; .
\end{align*}

\begin{figure}[tb]
\centering
\includegraphics[clip,height=6.0cm]{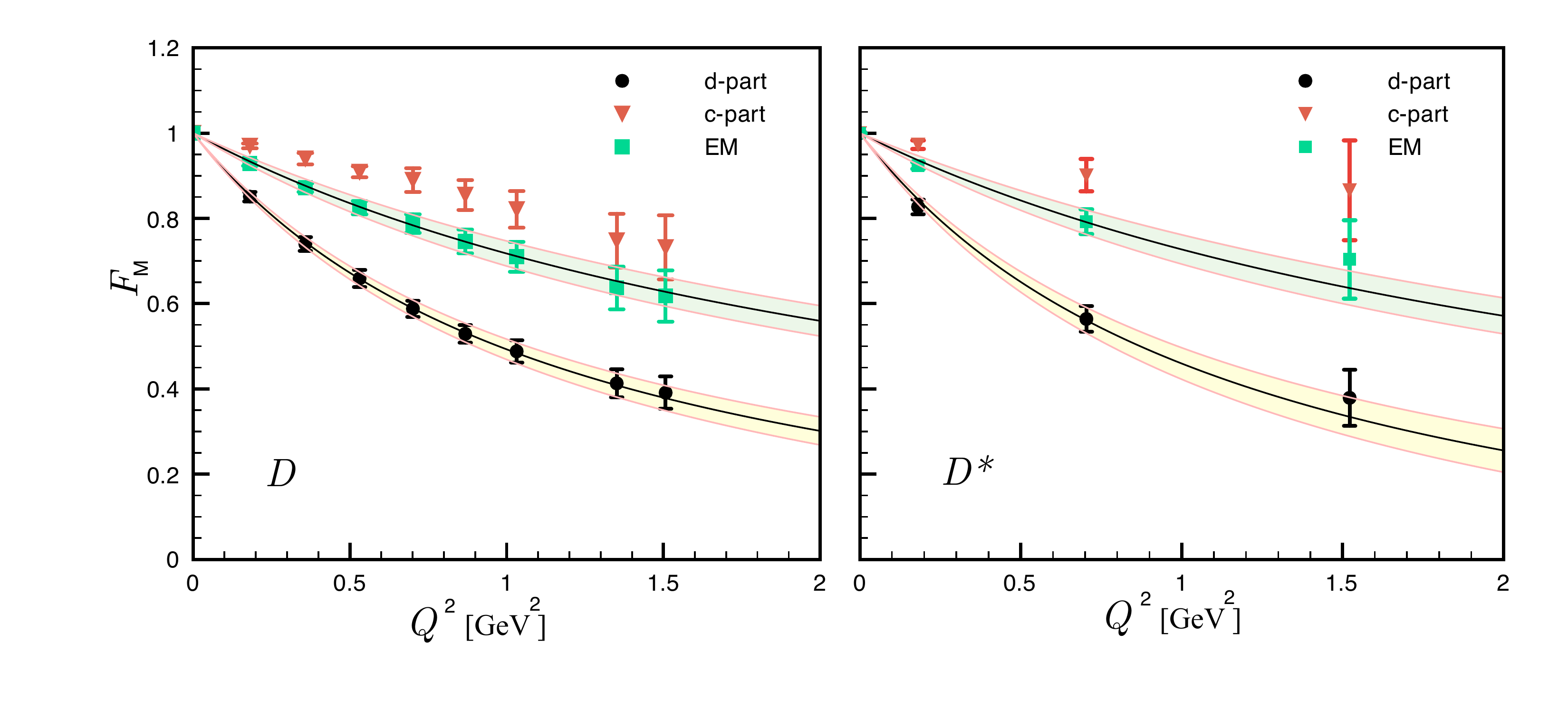}
\caption{Electromagnetic form factors of $D$ and $D^*$ as a function of $Q^2$ from \cite{Can:2012tx}.}
\label{fig:PLB719}
\end{figure}

One recent calculation \cite{Can:2012tx} uses 2+1 flavor gauge configurations with $a=0.0907(13)$ and  $m_\pi\in(300,410,570,700)$ MeV to extract $g_{DD^*\pi}$, $g_{DD\rho}$,
  $g_{D^*D^*\rho}$, the electromagnetic form factors and charge radii of $D$ and $D^*$ mesons. Figure \ref{fig:PLB719} shows the electromagnetic form factor as a function of the four-momentum transfer $Q^2$. Using a vector meson dominance approach with
\begin{align*}
F_V(Q^2)&=\left[1-\frac{Q^2}{m_\rho^2+Q^2}\frac{g_{D^{(*)}D^{(*)}\pi}}{g_\rho}\right]\; ,
\end{align*}

\begin{wrapfigure}[23]{r}{0.45\textwidth}
\centering
\includegraphics[clip,width=0.4\textwidth]{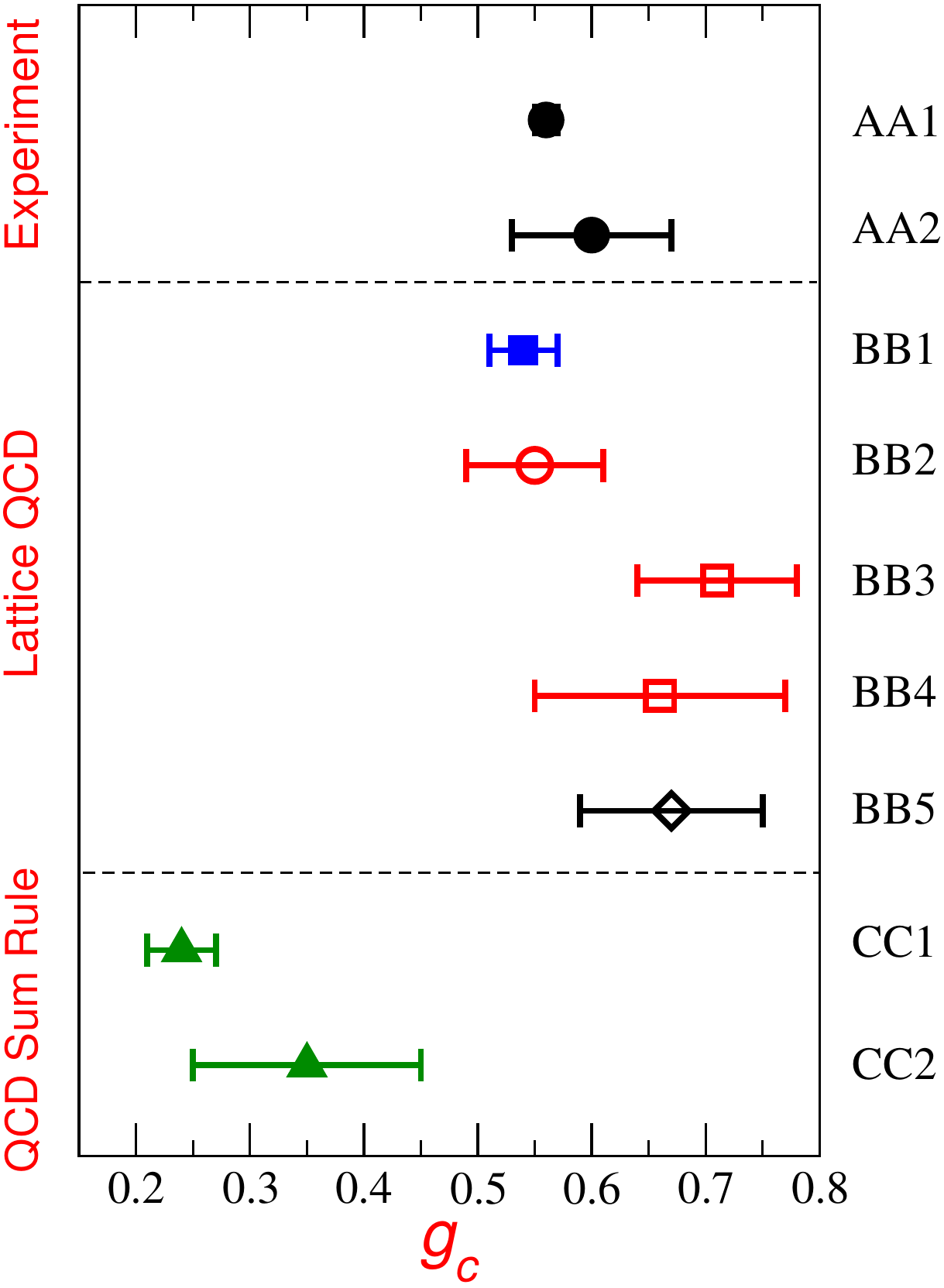}
\caption{Value of $g_c$, comparing experiment, Lattice QCD and Sum Rule calculations (from \cite{Becirevic:2012pf}). The blue square is the result from \cite{Becirevic:2012pf} and the red circle from \cite{Can:2012tx}.}
\label{fig:PLB721}
\end{wrapfigure}

\noindent the couplings
\begin{align*}
g_{DD\rho}&=4.84(34)\; ,\\
g_{D^*D^*\rho}&=5.94(56)
\end{align*}
are obtained. Notice that disconnected diagrams are neglected in this determination.

Another recent calculation \cite{Becirevic:2012pf} uses 2 flavor gauge configurations at 4 lattice spacings with
  $m_\pi\in(280,500)$MeV to determine $g_{DD^*\pi}$ at the physical point. The authors use nonperturbative renormalization and assess the systematic
  uncertainties in the chiral and continuum extrapolations.  Rather than quoting the result for $g_{DD^*\pi}$ directly, the authors report $g_c$ which is given by
\begin{align*}
g_{DD^*\pi}&=\frac{2\sqrt{m_Dm_{D^*}}}{f_\pi}g_c\; .
\end{align*}
Figure \ref{fig:PLB721} shows a compilation of recent lattice results along with the experimental values and results from QCD sum rules. The value from \cite{Becirevic:2012pf} (blue symbol and bar in the figure) leads to $\Gamma(D^{*+}\rightarrow\bar{D}^0\pi^+)=50\pm 5\pm 6$keV. 

\section{Searches for exotic charmed states}

\begin{figure}[htb]
\centering
\includegraphics[clip,height=5.0cm]{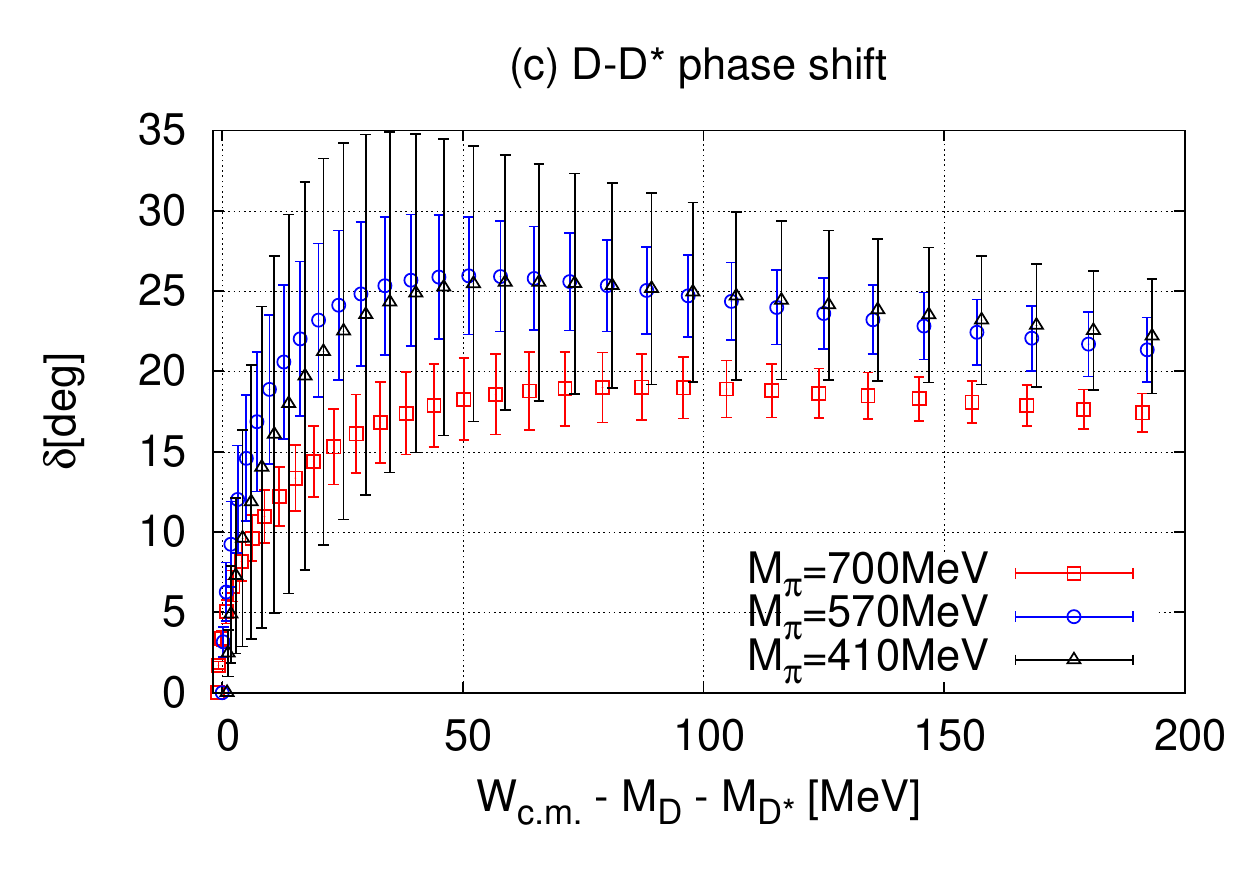}
\includegraphics[clip,height=5.0cm]{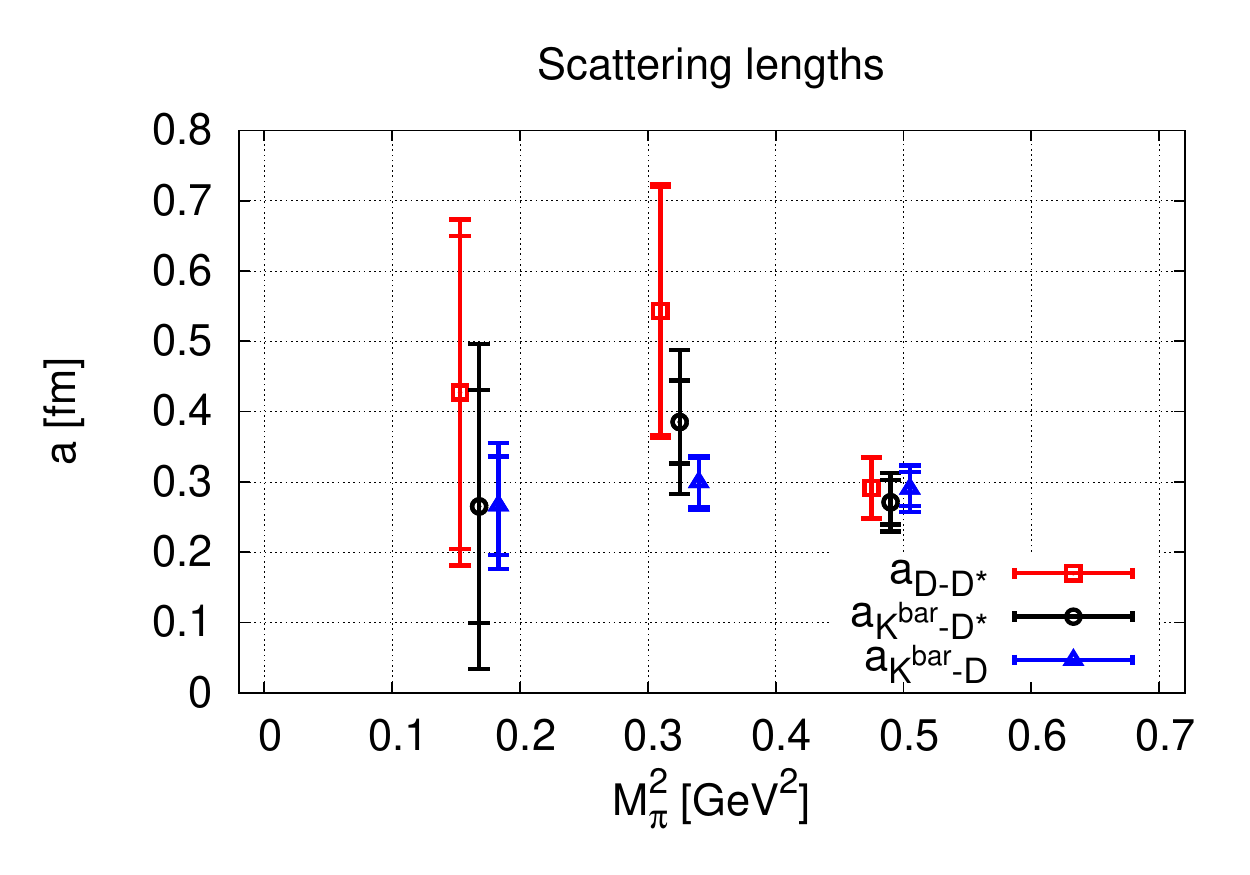}
\caption{Left: s-wave $I=0$ scattering phase shifts in the $DD^*$ channel. Right: Pion mass dependence of the scattering lengths. Both plots from \cite{Ikeda:2013vwa}.}
\label{fig:Ikeda}
\end{figure}

A recent study by the HALQCD collaboration investigates doubly-charmed and \linebreak charmed-strange tetraquarks \cite{Ikeda:2013vwa} on 2+1 flavor gauge configurations with $a=0.0907(13)$ fm and $m_\pi=410,570,700$ MeV, using the HALQCD method \cite{HALQCD:2012aa}. The search focuses on bound states or resonances in $DD$, $\bar{K}D$, $DD^*$ and $\bar{K}D^*$ interactions with flavor structure $cc\bar{u}\bar{d}$ and $cs\bar{u}\bar{d}$. These contain no quark line diagrams with quark annihilation. For the charm quarks a relativistic heavy quark action (a variant of the Fermilab method) is used. The HALQCD method consists of calculating a potential as a function of distance $r$ and solving the Schr\"odinger equation with given $V(r)$ to determine scattering phase shifts. An example for the resulting phase shifts is shown in the left pane of Figure \ref{fig:Ikeda}. The authors obtain a repulsive interaction in all $I=1$ channels and an attractive interaction in all $I=0$ channels considered. At the simulated $m_\pi$, no bound states or resonances are observed, although the authors state that the attraction becomes more prominent at light pion masses and that there is some indication that $BB^*$ with $I(J^P)=0(1^+)$ is bound \cite{Ikeda:2013vwa}.

\begin{wrapfigure}[17]{r}{0.57\textwidth}
\centering
\includegraphics[clip,width=0.57\textwidth]{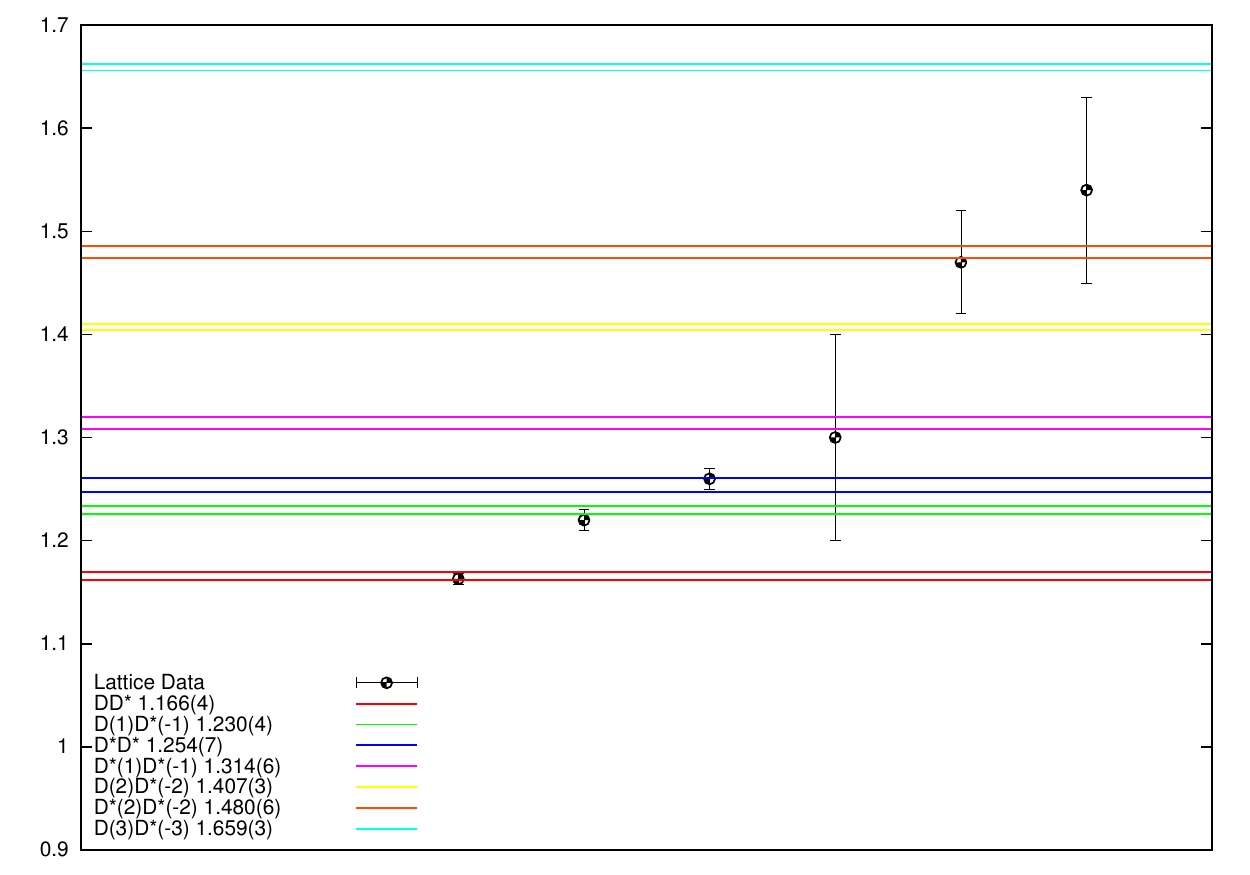}
\caption{Energy spectrum for quantum numbers $I(J^P)=0(1^+)$ from the preliminary results in \cite{Guerrieri:2014nxa}.}
\label{fig:Guerrieri}
\end{wrapfigure}
In \cite{Guerrieri:2014nxa} a preliminary search for doubly charmed tetraquarks is presented. Results are based on a 2 flavor simulation  with $a=0.075$fm and $m_\pi=490$MeV and lighter than physical charm quark mass. Tetraquarks with flavor structure $[cc][\bar{u}\bar{d}]$ and quantum numbers $I(J^P)=0(1^+), 1(1^+)$ are considered using a basis of tetraquark and meson-meson interpolators, including smeared interpolators. Figure \ref{fig:Guerrieri} shows some of the results. Beyond meson-meson states, no additional low-lying energy level is observed in this preliminary study.

\begin{figure}[htb]
\centering
\includegraphics[clip,height=4.5cm]{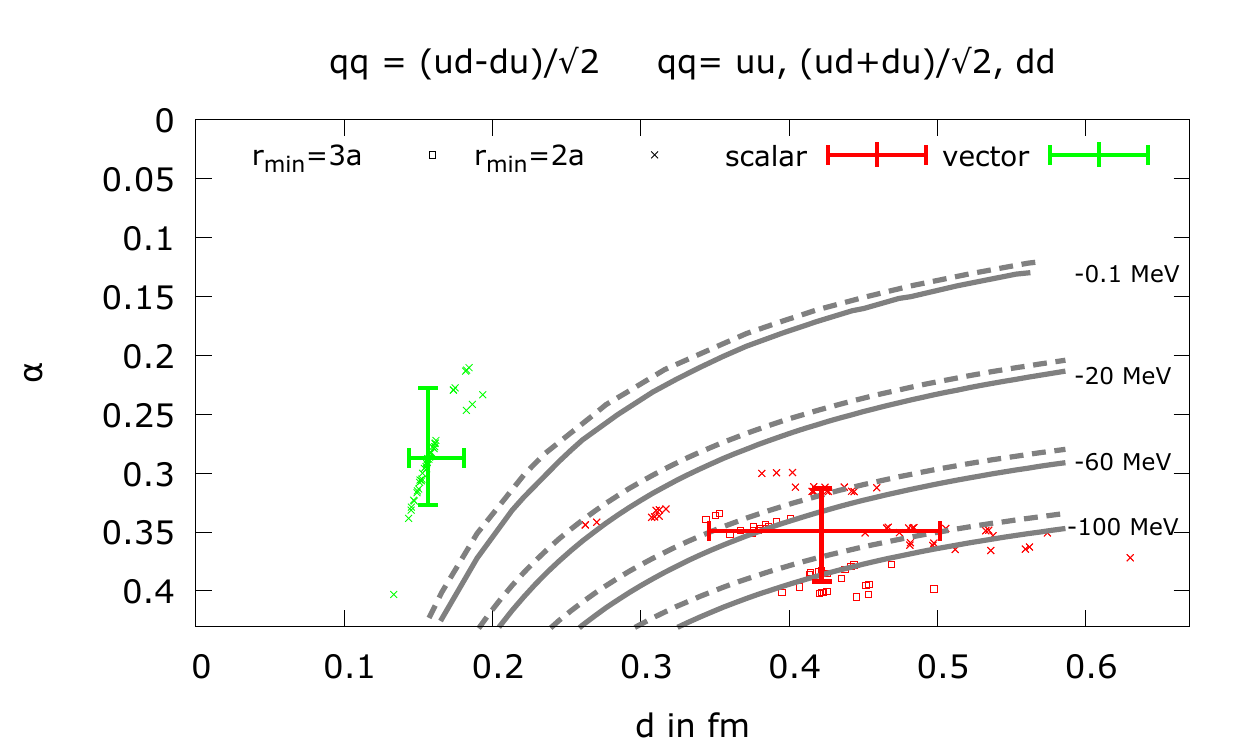}
\includegraphics[clip,height=4.5cm]{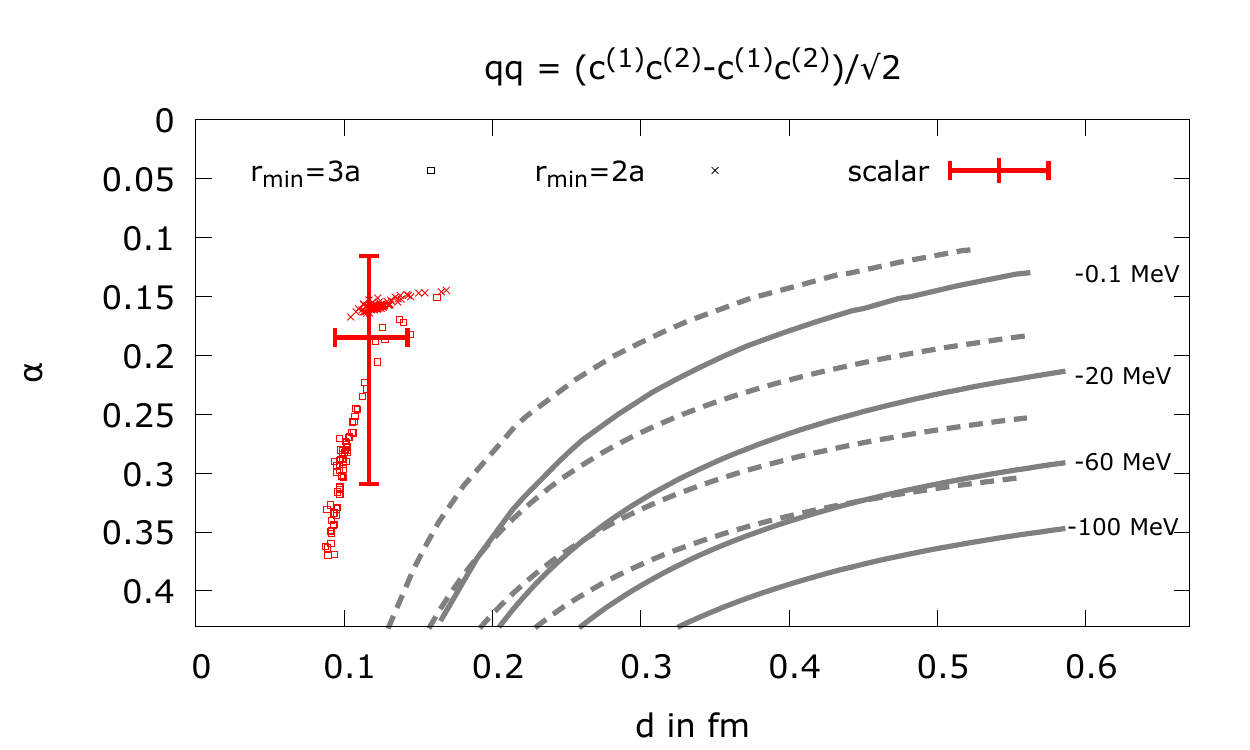}
\caption{Left: Binding energy isolines for two static and two light (u/d) quarks. Right: Same for two static and two charm quarks. Plots from \cite{Bicudo:2015vta}.}
\label{fig:Bicudo}
\end{figure}

Further studies search for tetraquarks in systems of two heavy and two light quarks with static heavy quarks \cite{Brown:2012tm,Bicudo:2012qt,Bicudo:2015vta}. Figure \ref{fig:Bicudo} shows results from a search for $ud\bar b\bar b$ $ss\bar b\bar b$ and $cc\bar b\bar b$ tetraquark bound states \cite{Bicudo:2015vta} obtained by studying potentials of two static antiquarks in the presence of two finite mass quarks. Lattices with $a=0.079$, $a=0.042$fm and $m_\pi\approx 350$MeV were used. The fit function used for the Lattice QCD potentials is given by
\begin{align*}
V(r)&=-\frac{\alpha}{r}\exp\left(-\left(\frac{r}{d}\right)^p\right) + V_0\; .
\end{align*}
In the system of two static and two light quarks a bound state is observed for the scalar channel, as shown in the left pane of Figure  \ref{fig:Bicudo}.

\section{Summary and conclusions}

Masses of $D_{(s)}^{(*)}$ ground states are well determined in current lattice simulations and recent results focus on their properties. For studies of excited states, a large number of energy levels can be extracted, including energy levels with a large overlap to hybrid meson interpolators \cite{Moir:2013ub}, providing further evidence for the existence of gluonic excitations. Until now very few simulations study close to threshold bound states and resonances, but first studies \cite{Mohler:2013rwa,Lang:2014yfa} suggest that the closeness of the $DK$ and $D^*K$ threshold is indeed very important for the $D_{s0}^*(2317)$ and $D_{s1}(2460)$ mesons, which was suggested a long time ago \cite{vanBeveren:2003kd} as a mechanism for explaining their unexpected mass and narrow width. Regarding scattering of D-mesons, preliminary results from the Hadron Spectrum Collaboration \cite{Moir:2013yfa} are very promising. Furthermore, there are results from several simulations of explicitly exotic tetraquark states \cite{Ikeda:2013vwa,Guerrieri:2014nxa,Brown:2012tm,Bicudo:2012qt,Bicudo:2015vta}. While these searches do not (yet)  obtain any evidence for charmed tetraquarks, some of the corresponding beauty states might exist \cite{Ikeda:2013vwa,Bicudo:2012qt,Bicudo:2015vta}. Beyond the current exploratory calculations, an urgent task for the lattice community will be better control of (heavy-quark) discretization effects by simulations on multiple lattice spacings.

\Acknowledgements
I would like to express my appreciation to my collaborators Christian Lang, Luka Leskovec, Sasa
Prelovsek and Richard Woloshyn. Furthermore I would like to thank Uktu Can, Graham Moir, Paula Perez-Rubio, Sasa Prelovsek for
providing me material for this talk. Fermilab is operated by Fermi Research Alliance, LLC under Contract No. De-AC02-07CH11359 with the United States Department of Energy.

\begin{small}
\begin{multicols}{2}

\end{multicols}
\end{small}

\end{document}